\documentclass[twocolumn,showpacs,preprintnumbers,amsmath,amssymb,prb,aps]{revtex4-1}
\usepackage{epsfig}
\usepackage{graphicx}
\usepackage{dcolumn}
\usepackage{bm}
\usepackage{textcomp}

\begin{document}

\title{Fabrication of setup for high temperature thermal conductivity measurement}
\author{Ashutosh Patel}
\altaffiliation{Electronic mail:ashutosh$\_$patel@students.iitmandi.ac.in}
\author{Sudhir K. Pandey}
\altaffiliation{Electronic mail:sudhir@iitmandi.ac.in}
\affiliation{School of Engineering, Indian Institute of Technology Mandi, Kamand 175005, Himachal Pradesh, India}
\date{\today}

\begin{abstract}
In this work, we report the fabrication of experimental setup for high temperature thermal conductivity ($\kappa$) measurement. It can characterize samples with various dimensions and shapes. Steady state based axial heat flow technique is used for $\kappa$ measurement. Heat loss is measured using parallel thermal conductance technique. Simple design, lightweight and small size sample holder is developed by using a thin heater and limited components. Low heat loss value is achieved by using very low thermal conductive insulator block with small cross sectional area. Power delivered to the heater is measured accurately by using 4-wire technique and for this, heater is developed with 4-wire. This setup is validated by using $Bi_{0.36}Sb_{1.45}Te_3$, polycrystalline bismuth, gadolinium, and alumina samples. The data obtained for these samples are found to be in good agreement with the reported data. The maximum deviation of 6 \% in the value $\kappa$ are observed. This maximum deviation is observed with the gadolinium sample. We also report the thermal conductivity of polycrystalline tellurium from 320 K to 550 K and the non monotonous behavior of $\kappa$ with temperature is observed. 
\end{abstract}

\maketitle

\section{Introduction}
In current scenario, finding new sources for power generation are very important to fulfill the demand of electricity. Various renewable energy sources like solar PV cell, wind turbine, etc. are discovered and currently are using at large scale. Various other sources have also been discovered but are in development stage. Power generation through thermolectric (TE) materials is one of them. Research is going on in this field to find good TE materials, specially for high temperature application. A good TE material should have a high Seebeck coefficient ($\alpha$), low thermal conductivity ($\kappa$), and high electrical conductivity ($\sigma$). Combining all three parameters, a common term introduced called figure of merit defined as $Z\overline{T} = \alpha^2\sigma \overline{T}/ \kappa$, where $\overline{T}$ is the mean temperature across TE material. A good TE material should have high Z$\overline{T}$ and for this, the Seebeck coefficient and electrical conductivity should be high, whereas the thermal conductivity should be low. The potential sites for thermoelectric generators are available at high temperature, so characterization of TE materials at high temperature are required. The measurement of all three parameters are required to find Z$\overline{T}$. Among all these three properties, $\kappa$ is most difficult to measure experimentally.

   Various available methods to measure $\kappa$ are discussed in Ref. [1]. Measurement methods are categorized in two parts, steady-state based and non-steady state based methods. Non steady state based methods like laser pulse method, 3$\omega$-method, etc. require very less time for measurement as compared to steady state based methods. Laser pulse method is used most widely in commercial instruments but it is limited to thin disc sample only.\cite{laserpulse} 3$\omega$-method is also a popular technique, where time delay between the heating of the sample and the temperature response is used to measure $\kappa$.\cite{3omega} Steady state based methods, like Axial heat flow technique, comparative technique, guarded heat flow technique, heat-flow meter technique, etc. are also used for measurement.\cite{method} The benefit of steady state based methods are that, setup cost is low and samples with various shapes and dimensions can be characterized. Steady state based axial heat flow method is a basic technique in which one-dimensional Fourier's law of thermal conduction equation is used to measure $\kappa$, which can be written as
   $$\kappa=\frac{\dot{Q}_s}{A}.\frac{l}{\Delta T}$$
Where $\dot{Q}_s$ is the net heat flow through the sample. A, $l$, and $\Delta T$ are cross-sectional area of the sample, its thickness, and temperature difference across it, respectively. The main difficulty with this method is to measure the heat flow through the sample ($\dot{Q}_s$) accurately. This difficulty arises due to the undefined amount of heat loss during the heat transfer process by conduction, convection and radiation, which becomes important at high temperature.\cite{amatya} Various instruments are reported based on this method for the measurement of $\kappa$. \textit{Muto et al.}\cite{muto} used a heat flux sensor to measure $\dot{Q}_s$. \textit{Amatya et al.}\cite{amatya} used comparative axial heat flow technique to measure $\dot{Q}_s$. In this technique, a sample is sandwiched between standard reference materials of known $\kappa$. Radiation losses are taken care by using some physical model. \textit{Zawilski et al.}\cite{zawilski} discussed a different approach to find the value of $\dot{Q}_s$, called parallel thermal conductance technique. In this technique, heat loss is measured by running the instrument without sample. This heat loss data is act as a baseline for $\kappa$ measurement. Using this technique, they performed measurement from 12 K to room temperature. Further, \textit{Dasgupta et al.}\cite{dasgupta} used this technique and performed measurement at high temperature. They developed a model in which they run the instrument at constant power without sample and with sample. The value of heat flow through the sample is obtained using heater constant and difference in equilibrium temperatures change in the heater block without sample and with sample. The value of heater constant is evaluated from the plot of equilibrium temperature change in the heater block without sample vs power supplied to the heater. Heater heats the side wall of copper rod and this copper block heat the sample. They put the sample holder in a glass chamber and kept it in a furnace.

 In the above discussed instruments, various techniques are used to measure $\dot{Q}_s$. Use of heat flux sensor limit the measurement temperature, as it is not suitable for high temperature applications. Similarly, comparative axial heat flow technique requires standard reference materials and number of temperature sensors. Radiation loss at high temperature is estimated by using some physical model. To get a more accurate result by using the method discussed by \textit{Zawilski et al.} heat loss should be as low as possible. Use of bulk heater causes more heat loss as it heat side walls and also high temperature surface is directly exposed to chamber environment. 

 In this work, we have developed a low cost setup for high temperature $\kappa$ measurement. The Parallel thermal conductance technique is used to measure heat flow through the sample. Heat loss is minimized by using low thermal conductive gypsum as an insulator block with small cross sectional area. Thin heater is built to heat the sample. Its negligible surfaces are exposed to vacuum environment, which also minimized the heat loss. The use of thin heater and small cross section insulator block minimizes the size of the sample holder assembly. This small size sample holder requires limited components, which make it lightweight. Thin heater is built with 4-probe contacts, by which we can measure accurate power delivered to the heater. This setup can characterize samples with various shapes and dimensions. $Bi_{0.36}Sb_{1.45}Te_3$, polycrystalline bismuth, gadolinium and alumina samples are used to validate the instrument. The data collected on these samples are found to be in good agreement with the reported data. We also report the $\kappa$ of the polycrystalline tellurium.
 
\begin{figure}
\includegraphics[width=9cm,height=6.7cm]{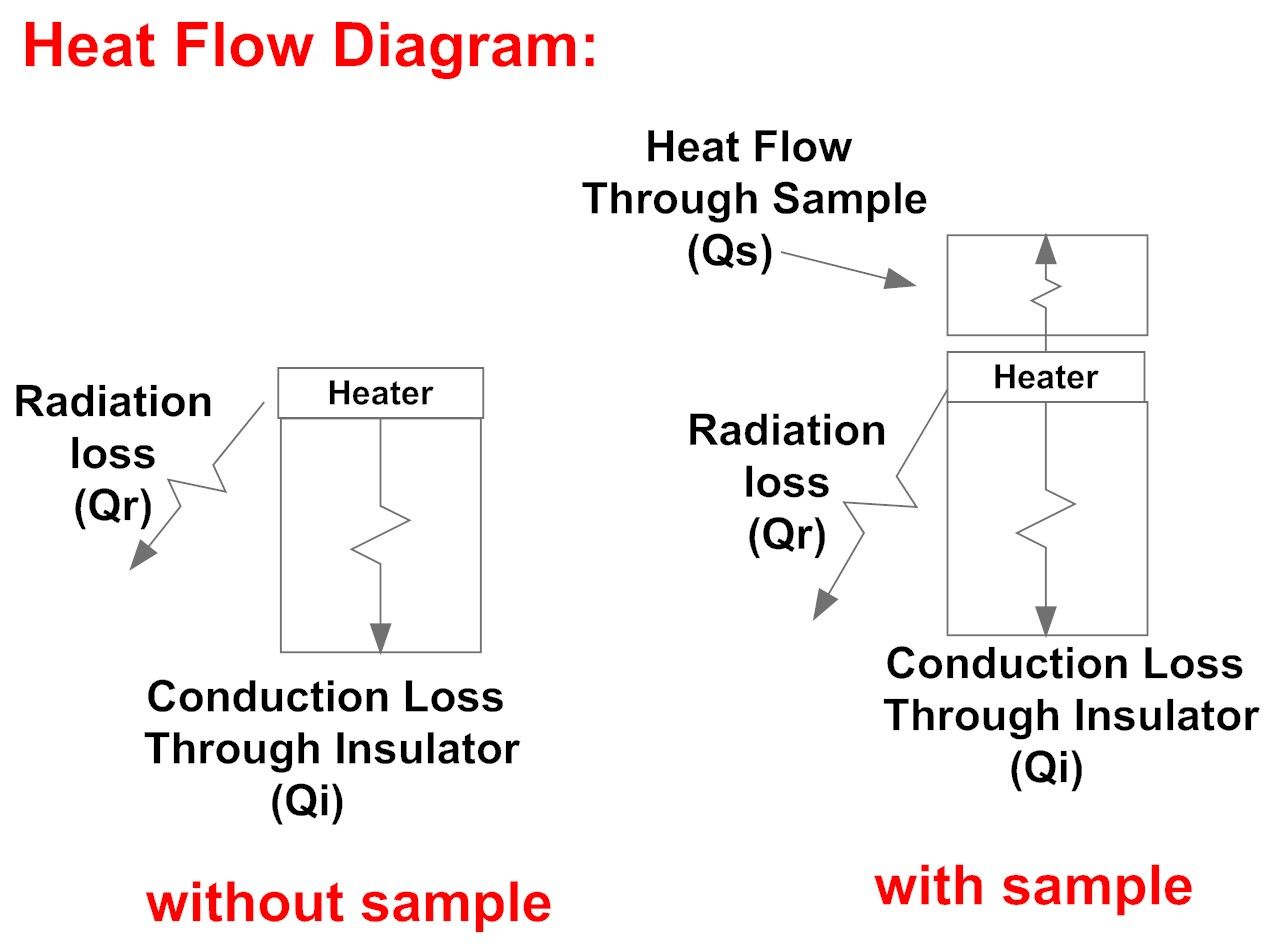}
\caption{Schematic diagram of heat flow without sample and with sample.}
\end{figure}

\section{MEASUREMENT METHODOLOGY}
One-dimensional Fourier's law of thermal conduction equation is used to measure the value of $\kappa$. As discussed in previous section, in this method accuracy of the $\kappa$ measurement mainly depends on the value of heat flow through the sample. It requires the estimation of heat loss by means of conduction, convection, and radiation.

 For the measurement setup described in this work, heat generated by the heater is mainly divided into two paths. One is through sample and another one is through the insulator block. Heat flow through the insulator block is evaluated by running the setup without sample. This method to find the heat loss is known as parallel thermal conductance technique.\cite{zawilski} The schematic diagrams of heat flow without sample and with sample are shown in Fig. 1. The heat loss data obtained by running the instrument without sample includes the conduction loss through the insulator block ($\dot{Q}_{ins}$) and also the radiation losses through the side walls of the insulator block and the surface of hot side copper block ($\dot{Q}_{rad}$). During heat loss measurement, hot side temperature with respect to vacuum chamber temperature ($T_{HR}$) is recorded along with the power supplied to the heater. As the power supply changes, $T_{HR}$ changes. The heat loss measurement is performed till the temperature for which $\kappa$ measurement has to be performed.
From Fig. 1, The heat loss equation can be written as    
$$\dot{Q}_\textit{l}=\dot{Q}_{ins}+\dot{Q}_{rad}$$
Where $\dot{Q}_\textit{l}$ is the net power delivered to the heater during heat loss measurement.

 During $\kappa$ measurement, the sample is placed in the holder. As shown in the Fig. 1, heat generated by the heater flows though insulator as well as sample also. The heat loss through side walls of the sample by means of conduction and convection to the air is less than 1 \% \ of the total heat flow through the sample, even at high hot side temperature (above 773 K).\cite{amatya} This heat loss can be ignored. The heat flow equation with a sample can be written as
$$\dot{Q}_{s} \approx \dot{Q}_{in}-\dot{Q}_\textit{l}$$
Where, $\dot{Q}_{in}$ is the net power delivered to the heater during $\kappa$ measurement. This $\dot{Q}_s$ obtained using above equation is used to calculate the $\kappa$ of the sample. In the above equation, equality sign is replaced by approximate symbol, as the value of heat loss through the sidewalls of the sample by means of radiation is ignored. This loss depends on the sample emissivity, geometrical shape and the temperature gradient across it. The whole measurement process are carried out at a constant room temperature (300 K).

\section{MEASUREMENT SETUP}

\begin{figure}[t]
\includegraphics[width=6cm,height=9.77cm]{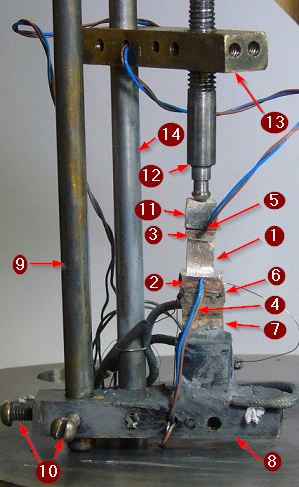}
\caption{Photograph of the sample holder.}
\end{figure}

The actual photograph of the sample holder assembly is shown in Fig. 2, where different components are represented by numbers. The sample, \textbf{\textit{1}}, is sandwiched between two copper blocks, \textbf{\textit{2}} \& \textbf{\textit{3}}, of $8mm \times 8mm$ cross section and 2 mm thickness. The two K-type PTFE shielded thermocouples, \textbf{\textit{4}} \& \textbf{\textit{5}}, of 36 swg are embedded at the center of each copper block. Measured $\Delta T$ includes the temperature gradient generated due to 1 mm of each copper block, the thermal resistance of interface surfaces and the sample. This instrument is used to measure low thermal conductivity (up to 30 W/m.K) sample, where the temperature gradient due to copper block is negligible as it has a very high thermal conductivity ($\sim$400 W/m.K at room temperature). The temperature gradient due to the interface surfaces has been minimized by using GaSn liquid metal eutectic, which has very low thermal contact parasitic (0.05 $Kcm^2/W$).\cite{amatya} High boiling temperature of GaSn liquid metal eutectic ($\sim$1800 K) make it suitable for high temperature application. 13 $\Omega$ thin heater, \textbf{\textit{6}}, is used to heat the sample. It is made by winding 40 swg kanthal wire over the alumina sheet of $7mm \times 7mm$ cross section and 0.6 mm of thickness. Each end of kanthal wire is welded with two nickel wires of 35 swg by using gas welding. At each end, one wire is used to supply power and another is used to measure the voltage across kanthal wire. This 4-wire technique measures exact power delivered to the heater. The current carrying nickel wires are kept shorter and exposed to the environment of vacuum chamber just after welded tip. Due to this, the heat generated in the nickel wires will not merge with heat generated through the heater. Also Adding two more nickel wires to achieve 4-wire configuration will not affect the heat loss value considerably, as these wires are very fine. This heater is further coated with high temperature cement to insulate it electrically and fixed over gypsum insulator block, \textbf{\textit{7}}, by using high temperature cement. Gypsum is used for this purpose as it has a very low thermal conductivity ($\sim$0.017 W/m.K at room temperature) and can be used at high temperature also.\cite{gypsum} As heat loss is very important and should be as low as possible, we take gypsum insulator block of $8mm \times 8mm$ cross section area and 25 mm thickness. Further, we minimized its effective cross sectional area by making a cylindrical hole across its thickness. This insulator block is supported by a brass plate, \textbf{\textit{8}}, by using high temperature cement. 

\begin{figure}[b]
\includegraphics[width=9cm,height=6.88cm]{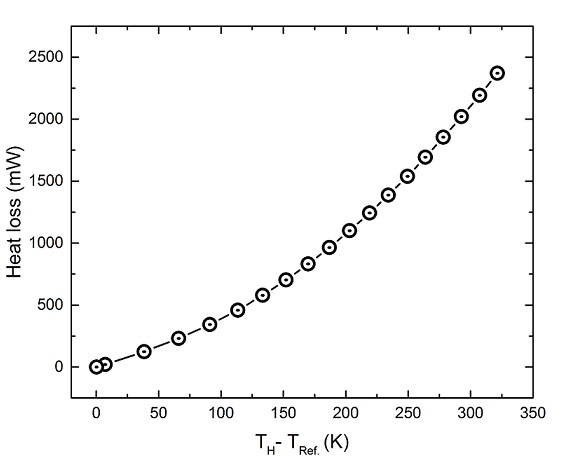}
\caption{Heat loss values at different $T_{HR}$.}
\end{figure}

\begin{figure*}[t]
\includegraphics[width=18cm,height=6.88cm]{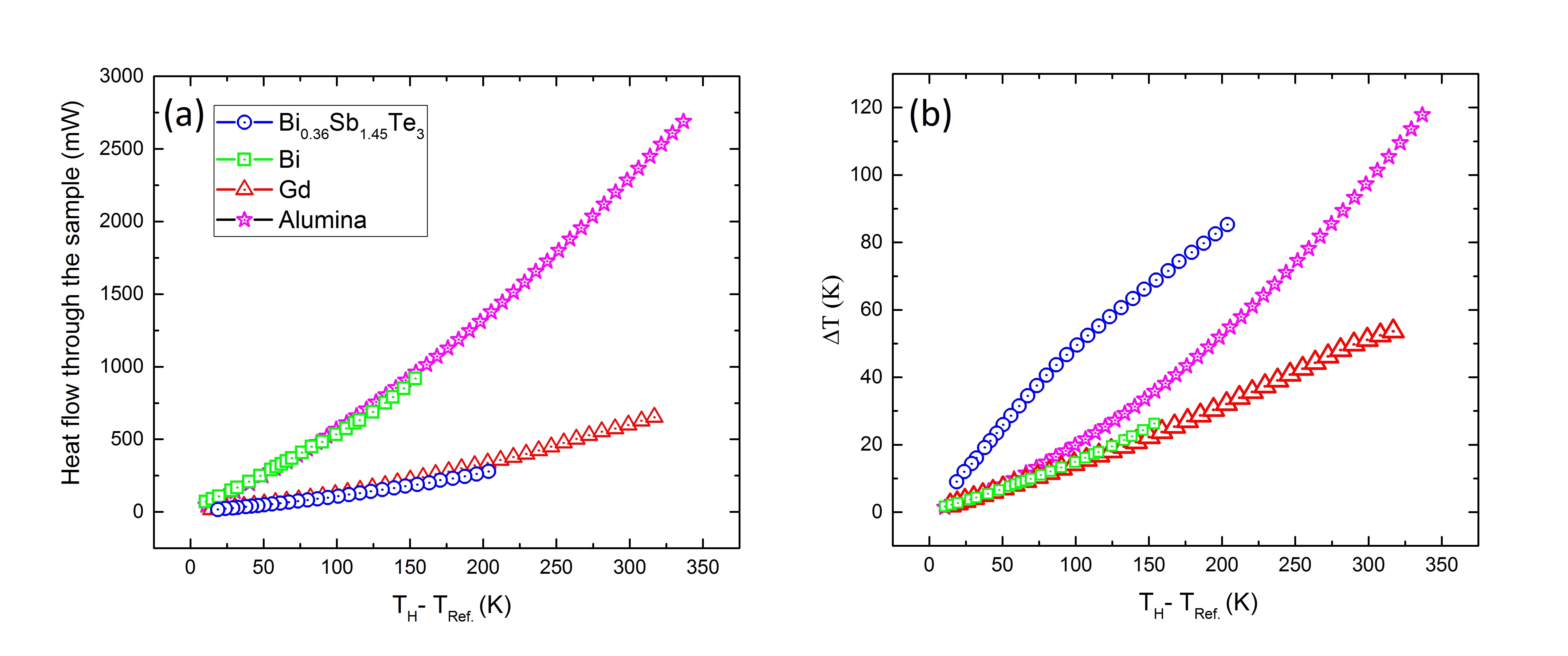}
\caption{(a) The values of heat flow and (b) Temperature gradient across the $Bi_{0.36}Sb_{1.45}Te_3$, polycrystalline bismuth, gadolinium, and alumina samples at different $T_{HR}$.}
\end{figure*}

This brass plate is fixed with a brass rod , \textbf{\textit{9}}, using screw , \textbf{\textit{10}}. Insulator block, \textbf{\textit{11}}, is also used at the cold side to insulate sample electrically and optimize $\Delta T$ across it. SS rod, \textbf{\textit{12}}, of 6 mm diameter and 60 mm length are used to apply pressure over the sample. This rod is having round tip at its end, which allows self aligning of the surfaces at the interface of sample and copper blocks. This self aligning ensures uniform pressure throughout the sample cross section. Another brass plate, \textbf{\textit{13}}, is used to screw this SS rod. It is fixed over a plane ss rod, \textbf{\textit{14}}, of 6 mm dia. Use of two different rods provide two independent paths for heat flow. Brass rod is for heat loss and ss rod is for heat flow through the sample. Brass rod and plane SS rod are fixed on a SS flange. On this SS flange, hermetically sealed electrical connector is fixed to make electrical connections. PT-100 RTD is connected to this connector to measure the temperature of the chamber. This sample holder assembly is placed inside the vacuum chamber, which is made by using seamless SS pipe of 10 cm diameter and 30 cm in height. KF25 port is provided over the vacuum chamber at the bottom. This port is used to connect this chamber with vacuum pump. Diffusion vacuum pump is used to create a vacuum inside the chamber upto a level of $\sim3\times10^{-5}$ mbar. At this level of vacuum, the conduction and convection loss through the air can be ignored because of this pressure thermal conductivity of air is 3$\times10^{-6}$ W/m.K . It is four order of magnitude lower than thermal conductivity at atmospheric pressure.\cite{4order}

 Sourcemeter is used to supply power to the heater and digital multimeter with scanner card is used to measure various signals. Sourcemeter is used in constant current mode and based on the resistance of the heater, power value is defined. LabVIEW based program is built to control the whole measurement process.

\section{RESULTS AND DISCUSSIONS}
We calibrated our instrument by using $Bi_{0.36}Sb_{1.45}Te_3$, polycrystalline bismuth, gadolinium, and alumina samples. Samples with various $\kappa$ values (varies from 1.4 W/m.K to 25 W/m.K) have been used to show the flexibility of the instrument. $Bi_{0.36}Sb_{1.45}Te_3$ sample is extracted from commercially available thermoelectric generator (TEC1-12706). The composition of the sample is obtained by performing EDX analysis. We also report the thermal conductivity of polycrystalline tellurium, which is not available to the best of our knowledge. The dimensions of the samples used for measurement are given in the table.

\newcolumntype{d}[1]{D{1}{\cdot}{#1} }
\begin{tabular}{l c c c c }
\hline
\hline
Sample&Cross section&Cross sectional&Thickness\\
name&type&area ($mm^2$)&($mm$)\\
\hline
$Bi_{0.36}Sb_{1.45}Te_3$&Rectangular&1.96&1.6\\
Poly. Bismuth&Rectangular&63&12\\
Gadolinium&Rectangular&2.56&2.4\\
Alumina&Rectangular&24&20\\
Poly. Tellurium&Rectangular&53&10\\
\hline
\hline
\end{tabular}
\\

 The reproducibility and repeatability of the instrument is verified by successive measurements on the same sample by remounting the sample for each measurement. The average value of $\kappa$ along with error bar are shown for all the samples.

\subsection{Heat loss, heat flow, and temperature gradient measurements}

\begin{figure*}[t]
\includegraphics[width=18cm,height=14.6cm]{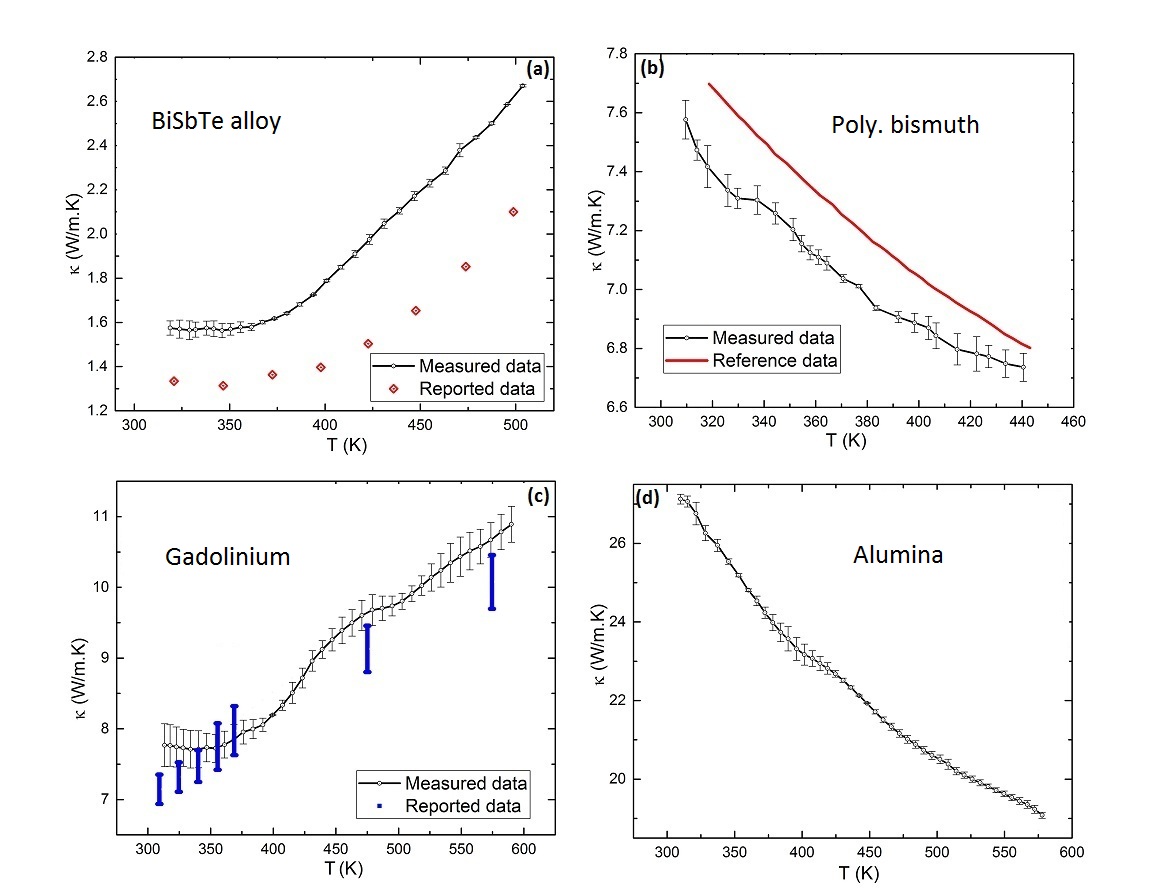}
\caption{(a) Thermal conductivity of $Bi_{0.36}Sb_{1.45}Te_3$ at different $\overline{T}$, (b) Thermal conductivity of polycrystalline bismuth at different $\overline{T}$, (c)  Thermal conductivity of gadolinium at different $\overline{T}$ , (d) Thermal conductivity of alumina at different $\overline{T}$.}
\end{figure*}

Heat loss values are very important for $\kappa$ measurement. At the start of the measurement, the value of $\dot{Q}_{s}$ is very low.  A small error in its value can change the result by considerable magnitude. For better result, $\dot{Q}_\textit{l}$ should be as low as possible. It is achieved by using low thermal conductive gypsum as insulator block. Again, the heat loss value is further minimized by reducing the effective surface area. A plot of heat loss with respect to $T_{HR}$ is shown in Fig. 3. This heat loss data follow cubic degree polynomial, which is obtained by fitting of heat loss data with respect to $T_{HR}$. Initially, $\sim$ 3 mW/K of heat loss is observed. This rate is increasing with increase in temperature, which is due to the increase in thermal conductivity of gypsum and radiation losses.

 The value of heat flow through the samples are evaluated by subtracting the heat loss data. The obtained values of $\dot{Q}_{s}$ for the $Bi_{0.36}Sb_{1.45}Te_3$, polycrystalline bismuth, gadolinium and alumina samples with respect to $T_{HR}$ are shown in the Fig. 4(a). The closer values of $\dot{Q}_{s}$ with $T_{HR}$ for $Bi_{0.36}Sb_{1.45}Te_3$ and gadolinium samples are observed. This $\dot{Q}_{s}$ depends on the total thermal resistance of the circuit, which include sample, cold side insulator block, SS rod, brass plate and brass rod. So, sample having low or high thermal resistance does not change the total thermal resistance by large value, which results almost same heat flow. The $\dot{Q}_{s}$, sample dimension and value of $\kappa$ decides the $\Delta T$ across it. Due to large cross sectional area of bismuth and high $\kappa$ of alumina, the values of $\Delta T$ with these samples are very small ($<$1 K) at the start of the measurement. This very small value of $\Delta T$ leads to give large uncertainty, because it also includes very small temperature gradient due to the interface surfaces and copper blocks which we have ignored. To nullify these gradients, the measured value of $\Delta T$ should not be too small. This issue is resolved by removing the insulator block used above the cold side copper block. This minimizes the thermal resistance of the circuit and results in large increment of $\dot{Q}_{s}$. This large increment in $\dot{Q}_{s}$ increases the $\Delta T$ proportionally. The values of $\Delta T$ with respect to $T_{HR}$ for all the four samples are shown in Fig. 4.(b). At the start of the measurement, the value of $\Delta T$ for $Bi_{0.36}Sb_{1.45}Te_3$, polycrystalline bismuth, gadolinium, and alumina samples are 9 K, 1.7 K, 2 K, and 1.4 K, respectively. These values increase with the increase in temperature and reached to 90 K, 26 K, 42 K, and 120 K, respectively.

\subsection{Thermal conductivity measurement}

Measured and reported values of $\kappa$ for $Bi_{0.36}Sb_{1.45}Te_3$, polycrystalline bismuth, and gadolinium samples at various temperatures are shown in Fig. 5 (a), 5 (b), and 5 (c), respectively. Measured values of $\kappa$ for alumina sample at various temperatures are shown in Fig. 5 (d). 

 Initially, the value of $\kappa$ for $Bi_{0.36}Sb_{1.45}Te_3$ sample is 1.57 W/m.K. It decreases with increase in temperature with very small rate till $\overline{T}$=350 K and reaches to 1.565 W/m.K. After this temperature, $\kappa$ increases with temperature and reaches to 2.67 W/m.K at $\overline{T}$=502 K.
\textit{Ma et al.}\cite{bisbte} reported $\kappa$, in which they used commercial $BiSbTe$ ingot as a sample. They obtained composition of sample by comparing XRD data of the sample with available XRD data of $Bi_{0.5}Sb_{1.5}Te_3$. Our data show similar behavior compared to the reported data. From 320 K to 375 K, almost constant deviation of 0.22 W/m.K is observed. It increases with increase in temperature and reach to 0.5 W/m.K at 500 K.
From the available experimental data in Ref. [11], it is observed that the value of $\kappa$ increases with increase in the concentration of bismuth in BiSbTe alloy. The concentration of bismuth in our sample is more compared to the sample used in the reference, which may be the reason that the value of $\kappa$ for our sample is higher than the reported value of $\kappa$. 

In Bismuth, the electron and phonon components of $\kappa$ are comparable in magnitude above 120 K. The value of $\kappa$ decreases with temperature due to decrease in the magnitude of electron and phonon components with temperature.\cite{bismuthphonon} For a polycrystalline bismuth sample, we observed similar behavior. The value of $\kappa$ at $\overline{T}$=318 K is 7.57 W/m.K. It decreases with increase in temperature and reaches to 6.73 W/m.K at $\overline{T}$=440 K. Our data shows similar behavior with the data available in handbook.\cite{bismuth} A deviation of 0.3 W/m.K is observed  at $\overline{T}$=318 K. This deviation decreases with temperature and reach to 0.08 W/m.K. The maximum deviation of 0.3 W/m.K is observed at 318 K, which is 4 \% of the value of the $\kappa$.
 
 Gadolinium is a rare earth metal, which shows increasing behavior in thermal conductivity with temperature after ferromagnetic-antiferromagnetic transition. Its ferromagnetic-antiferromagnetic transition occurs near room temperature.\cite{rare} Our data of gadolinium sample measured from 315 K to 590 K shows increasing behavior. Its thermal conductivity at $\overline{T}$=315 K is 7.76 W/m.K, with temperature $\kappa$ increases and at at $\overline{T}$=590 K, its value is 10.9 W/m.K.
Data of gadolinium sample are compared with the reported data available in Ref. [15]. In this reference, values of $\kappa$ are obtained by using thermal diffusivity data. A deviation of 0.37 W/m.K is observed at 315 K. This deviation decreases with temperature and match closely with reference data at 340 K. At 475 K and 575 K, deviation of $\sim$0.18 W/m.K is observed. The maximum deviation of 0.37 W/m.K is observed at 315 K, which is the 5 \% of the reported value of the $\kappa$.

 It is observed that the thermal conductivity of oxide material decreases with temperature.\cite{oxide} Our data for alumina sample show similar behavior. At $\overline{T}$=310 K, the value of $\kappa$ is 27.1 W/m.K. It decreases with increase in temperature and reaches to 19 W/m.K at $\overline{T}$=578 K. 
For alumina sample, the values of $\kappa$ from various sources are compared in Ref. [17]. Near room temperature its value lies in the range of 24 W/m.K and 30 W/m.K. It decreases with temperature and at 600 K its value lies between 13.5 W/m.K to 17.5 W/m.K. Our data show similar behavior compared to the reported data. At 310 K, our data lies in the range of the reported data and at 578 K, a deviation of 0.5 W/m.K is observed.

\begin{figure}[t]
\includegraphics[width=9cm,height=6.88cm]{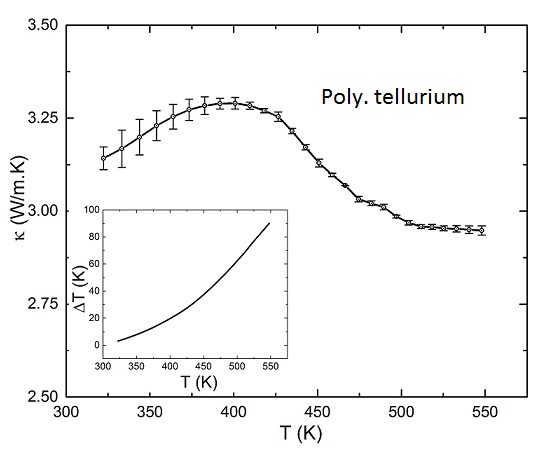}
\caption{Thermal conductivity of polycrystalline tellurium at different $\overline{T}$ (variation in $\Delta T$ with $\overline{T}$ is shown in the inset).}
\end{figure}

 We also performed measurement over polycrystalline tellurium sample. It is taken from commercially available ingot. The values of $\kappa$ for this sample with respect to $\overline{T}$ are shown in Fig. 5. Variation in $\Delta T$ along with $\overline{T}$ is shown in the inset of Fig. 6. At $\overline{T}$=320 K, $\Delta T$ is 3 K. $\Delta T$ increases and reaches to 90 K at $\overline{T}$=550 K. At $\overline{T}$=320 K, $\kappa$ is 3.14 W/m.K. It increases with increase in temperature with a very small rate till $\overline{T}$=400 K, where its value is 3.3 W/m.K. After this temperature, its value decreases and reached to 2.95 W/m.K  at $\overline{T}$=550 K.

 Thermoelectric properties of polycrystalline tellurium with different charge carrier concentrations are studied and measured with  by \textit{Lin et al.}\cite{tellurium}. Sample is prepared by melting and followed by quenching in cold water and annealing. charge carrier concentration tuned by using dopants including phosphorus (P), arsenic (As), antimony (Sb) and bismuth (Bi). With different carrier concentration, the value of $\kappa$ changes. Its maximum value near room temperature is $\sim$2.25 W/m.K for highest carrier concentration and with temperature it decreases.

 A wide variety of samples are used for measurement, whose $\kappa$ value lies in the range of 1.5 W/m.K to 27 W/m.K. At high temperature, previously only \textit{Dasgupta et al.}\cite{dasgupta} used parallel conductance technique for $\kappa$ measurement. They performed high temperature measurement with samples only upto a value of 2.5 W/m.K. By comparing Fig. 4 (b) with Fig. 5, the behavior of $\kappa$ for all four samples shown in Fig. 5 can be also obtained, by obtaining change in the slope of $\Delta T$ vs temperature graph (Fig. 4 (b)). The increase in the slope of graph shows decreasing behavior of $\kappa$, while decrease in the slope shows increasing behavior of $\kappa$. We found reproducibility of our data with a maximum error bar of $\pm$3 \%. We also observed a maximum deviation of 6 \% for gadolinium sample compared to the reported data at high temperature. Since we have ignored the radiation loss through the sidewalls of the sample in the heat flow measurement, may be the possible reason for the observed deviation.

\section{CONCLUSION}
In this work, we have developed simple, low cost and user friendly setup for high temperature thermal conductivity measurement. The wide verity of samples with various shapes and dimensions can be characterize using this setup. Good reproducibility of measured data is observed with a maximum error bar of $\pm$3 \%. The parallel thermal conductance technique is used for measurement and data measured with better accuracy than previously reported literature using this method. We measured $\kappa$ at high temperature using samples upto a value of 27 W/m.K. This is achieved by low heat loss and accurate power measurement by using 4-wire technique. Thin heater, simple design and limited components, make this setup light weight and small in size. This setup is validated by using $Bi_{0.36}Sb_{1.45}Te_3$, polycrystalline bismuth, gadolinium, and alumina samples. The measured data were found in good agreement with the reported data and observed a maximum deviation of 6 \% with the gadolinium sample, which indicate that this instrument is capable to measure $\kappa$ with fairly good accuracy. We also reported the thermal conductivity of polycrystalline tellurium and observed the non monotonous behavior of $\kappa$ with temperature. 

\section{ACKNOWLEDGEMENTS}
The authors acknowledge R S Raghav and other workshop staffs for their support in the fabrication process of vacuum chamber and sample holder parts.

\end{document}